\documentclass[aps,prb,twocolumn,showpacs]{revtex4}
\usepackage{graphicx}
\usepackage{latexsym}
\usepackage{amsmath}
\usepackage{graphics}
\usepackage{amssymb}
\usepackage{layout}
\usepackage{verbatim}
\usepackage{amsfonts,epsfig}
\newcommand{\ket}[1]{\left|#1\right>}
\newcommand{\bra}[1]{\left< #1 \right|}

\begin{document}
\title{Kinetics of the superconducting charge qubit in the presence of a quasiparticle}

\author{R. M. Lutchyn$^{1}$, L. I. Glazman$^{1}$, and A. I. Larkin$^{1,2}$}

\affiliation{$^{1}$ W.I.\ Fine Theoretical Physics Institute,
University
of Minnesota, Minneapolis, Minnesota 55455, USA \\
$^{2}$ L. D. Landau Institute for Theoretical Physics, Moscow,
117940, Russia}
\date{\today }
\begin{abstract}
We investigate the energy and phase relaxation of a
superconducting
  qubit caused by a single quasiparticle. In our model, the qubit is
  an isolated system consisting of a small island (Cooper-pair box)
  and a larger superconductor (reservoir) connected with each other by
  a tunable Josephson junction. If such a system contains an odd number
  of electrons, then even at lowest temperatures a single
  quasiparticle is present in the qubit. Tunneling
  of a quasiparticle between the reservoir and the
Cooper-pair box results in the relaxation of the
  qubit.  We derive master equations governing the evolution of the
  qubit coherences and populations. We find that the kinetics of the
  qubit can be characterized by
  two time scales - quasiparticle escape time from the reservoir to the
  box $\Gamma_{\textrm{in}}^{-1}$ and quasiparticle relaxation time
  $\tau$. The former is determined by the dimensionless normal-state
  conductance $g_{_T}$ of the Josephson junction and one-electron
  level spacing $\delta_r$ in the reservoir
  ($\Gamma_{\textrm{in}}\!\sim\!g_{_T}\delta_r$), and the latter is
  due to the electron-phonon interaction. We find that phase coherence is
  damped on the time scale of $\Gamma_{\textrm{in}}^{-1}$.  The qubit
  energy relaxation depends on the ratio of the two characteristic
  times $\tau$ and $\Gamma_{\textrm{in}}^{-1}$ and also on the ratio
  of temperature $T$ to the Josephson energy $E_{_J}$.
\end{abstract}
\maketitle
\section{Introduction}
Recent experiments with superconducting charge qubits demonstrated
coherent oscillations between two charge states of a
superconducting island, a so-called Cooper-pair box (CPB), in a
single-electron device~\cite{Nakamura, Wallraff, Vion, Blais}. A
device with a large superconducting gap $\Delta> E_c > E_{_J} \gg
T$ can be controlled with the gate voltage and magnetic flux, and
has only one discrete degree of freedom: the number of Cooper
pairs in the box. (Here $E_c$ is the charging energy of the
island, and $E_{_J}$ is an effective Josephson energy of its
junctions with the reservoir; $E_{_J}$ can be tuned by the flux.)
The practical implementation of superconducting qubits requires
long coherence times~\cite{DiVincenzo}. Although the contribution
of the quasiparticles to the decoherence of the existing qubits is
not the leading one~\cite{Devoret, Astafiev}, it limits qubit
operations on the fundamental level. This motivates us to study
the qubit dynamics in the presence of a quasiparticle.

In this paper we consider energy and phase relaxation in a
superconducting charge qubit due to the presence of a single
quasiparticle. Passing of a quasiparticle through the Josephson
junction leads to the escape of the qubit out of a two-level
system Hilbert space, and thus determines the decay rate of
coherent oscillations. In a previous paper~\cite{Lutchyn}, we
evaluated rates of the elementary acts involving quasiparticle
tunneling. In particular, we showed that the rate of tunneling
into the CPB $\Gamma_{\textrm{in}}$ is determined by the
dimensionless (in units of $e^2/h$) conductance $g_{_T}$ of the
junction between the CPB and the reservoir, and level spacing in
the reservoir $\Gamma_{\textrm{in}}\sim g_{_T}\delta_r/4\pi$. The
rate of tunneling out of CPB is $\Gamma_{\textrm{out}}\sim
g_{_T}\delta_b/4\pi$ with $\delta_b$ being level spacing in the
box. Taking into account the difference in the volumes ($V_r \gg
V_b$), one can notice that $\Gamma_{\textrm{out}} \gg
\Gamma_{\textrm{in}}$. Therefore, for a sufficiently small box the
quasiparticle dwelling time in the CPB is very short, and the
qubit spends most of the time in the ``good" part of the Hilbert
space. Nevertheless, the evolution of the qubit will be affected
by quick ``detours'' the qubit takes outside that part of the
Hilbert space. We demonstrate that even a single detour destroys
the coherence of the qubit. Combined with the phonon-induced
relaxation of the non-equilibrium quasiparticle in the reservoir,
the detours also lead to the relaxation of the populations of the
qubit states. We derive and solve the master equations for the
dynamics of the qubit, which describe its relaxation caused by a
quasiparticle.

The paper is organized as follows. We begin in Sec.~\ref{quali}
with a qualitative discussion and brief overview of the main
results. In Sec.~\ref{derivation} we derive the microscopic master
equations for the dynamics of the qubit without quasiparticle
relaxation. We solve these equations and discuss the kinetics of
the coherences and populations in the Sec. \ref{seccoher} and
\ref{norel}. In Sec.~\ref{withrelax} we incorporate mechanisms of
quasiparticle relaxation into the master equations and solve them
for fast and slow quasiparticle relaxation limits. Finally, in
Sec.~\ref{conclusion} we summarize our main results.

\section{Qualitative Considerations and Main Results}\label{quali}
Dynamics of the superconducting charge qubit is  conventionally
described by an effective Hamiltonian~\cite{Shnirman}
\begin{equation}\label{Hqubit}
H_{\textrm{qb}}=E_c(N-N_g)^2+H_{_J},
\end{equation}
where $E_c$ is charging energy of the box, $N$ is the charge of
  the CPB in units of one-electron charge $e$, and $N_g$ is the
dimensionless gate voltage. The Hamiltonian of  Cooper-pair
  tunneling $H_{_J}$ is defined as
$H_{_J}=-\frac{E_{_J}}{2}\left(\ket{N+2}\bra{N}+H.c.\right)$,
where $E_{_J}$ is the Josephson energy. In the case of a large
  superconducting gap
\[\Delta> E_c> E_{_J} \gg T,\]
the existence of quasiparticles is usually neglected, and the
dynamics of the system is described by Hamiltonian (\ref{Hqubit})
with only one discrete degree of freedom, the excess number of
Cooper pairs in the box. The qubit can first be  prepared in a
state with an integer number of pairs in the CPB, {\it e.g.}, in
state $\ket{N}$ with even $N$, and then tuned to the operating
point by adjusting the gate voltage to the value $N_g=N+1$. The
charge degeneracy between states $\ket{N}$ and $\ket{N+2}$ at this
point is lifted by the Josephson tunneling, and the states of the
qubit are described by the symmetric and antisymmetric
superposition of the charge states,
$\ket{-}=\frac{\ket{N}+\ket{N+2}}{\sqrt{2}}$ and
$\ket{+}=\frac{\ket{N}-\ket{N+2}}{\sqrt{2}}$ with energies
\begin{eqnarray}\label{omega}
\omega_{-}=E_c-E_{_J}/2 \mbox{ and } \omega_{+}=E_c+E_{_J}/2,
\end{eqnarray}
respectively.  Other charge states have much higher energy, and
effectively the Cooper-pair box reduces to a two-level system.
Being coherently excited by such tuning, the qubit oscillates
  between the states $\ket{+}$ and $\ket{-}$ with the frequency
  defined by the Josephson energy $E_{_J}$.

 The presence of a quasiparticle with a continuum excitation
  spectrum provides a channel for relaxation of the qubit. If the
  state $\ket{N}$ is prepared in equilibrium conditions, then the
  quasiparticle resides in the reservoir part~\cite{Lutchyn} of the
  qubit. Upon tuning of the qubit from state $\ket{N}$ to the
  operating point, a charge degeneracy point for the system is passed
  at $N_g=N+1/2$, see Fig. 1. (Hereafter we assume equal
  superconducting gap energies in the reservoir and Cooper-pair box.)
  However, if tuning is performed fast enough, the quasiparticle
  remains in the reservoir~\cite{Guillaume}.

\begin{figure}
\centering
\includegraphics[width=3.0in, scale=1]{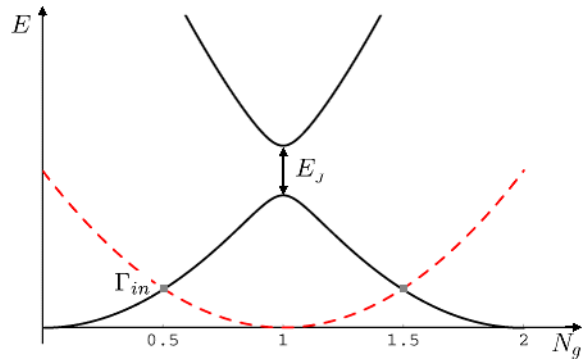}
\caption{(color online). Energy of the Cooper-pair box as a
  function of dimensionless gate charge $N_g$ in units of $e$ (solid
  line). Near the degeneracy point ($N_g=1$) Josephson coupling mixes
  charge states and modifies the energy of the CPB. The dashed line
  corresponds to the charging energy of the CPB with an unpaired electron in
  the box.  At $N_g=0.5$, the tunneling rate $\Gamma_{\textrm{in}}$ lifts
  the degeneracy between the ground state of the CPB (solid line) and a
  state with a single quasiparticle in CPB (dashed line). We
  assume equal superconducting gap energies in the reservoir and CPB.} \label{fig0}
 \end{figure}

 The coherent charge oscillations at the operating point of the qubit
 continue until the particle finds its way into the CPB. On average,
 this occurs on a time scale of the order of
 $(g_{_T}\delta_r/4\pi)^{-1}$.  There are several assumptions that
 allow for this estimate~\cite{Lutchyn}. First, the quasiparticle
 level spacings $\delta_b$ and $\delta_r$ in the CPB and reservoir,
 respectively, must be small compared with the temperature $T$, which
 defines the initial width of the energy distribution of the
 quasiparticle.  Second, the fluctuations of the potential between the
 grains must exceed $\delta_r$, see, {\it e.g.},
 Ref.~[\onlinecite{Blanter}]. Third, we neglected the difference between
 $\Delta$ and $E_c$ when including in the estimate the density of
 states and tunneling matrix elements of a quasiparticle at energy $\sim E_c$ above the gap in the CPB.
 Under these conditions, the average time it takes the quasiparticle
 to leave the reservoir and enter the CPB is of the order of the inverse level
 width of a state in the reservoir with respect to leaving it through
 the junction of conductance $g_{_T}$
\begin{equation}
\Gamma_{\textrm{in}}^{-1}\sim
\left(\frac{g_{_T}\delta_r}{4\pi}\right)^{-1}. \label{gammain}
\end{equation}

Once the quasiparticle enters the CPB, the charging energy that
the qubit has at operation point is transformed into the kinetic
energy of the quasiparticle, see Fig.~\ref{fig0}.  The
quasiparticle may escape the CPB leaving the qubit in the excited
or ground state, see Fig. 2. The rates of escape into these states
are different due to the difference of the kinetic energies
available to the quasiparticle upon the escape and due to the
energy dependence of the superconducting density of states $\nu
(\varepsilon)$.  If the qubit ends up in the excited state upon
the escape, then only energy $\varepsilon\sim T$ is available for
the quasiparticle, and $\nu (T)\sim\delta_r^{-1}(\Delta/T)^{1/2}$
(we used here the condition $\Delta\gg T$). The corresponding
escape rate is
\begin{equation}
\Gamma_{\textrm{out}}^{_{\ket{+}}}\sim
\frac{g_{_T}\delta_b}{4\pi}\sqrt{\frac{\Delta}{T}}.
\label{gammaplus}
\end{equation}
If the qubit arrives in the ground state, then energy $\sim
E_{_J}$ is available to the quasiparticle, and its density of
states in the final state is $\nu
(E_{_J})\sim\delta_r^{-1}(\Delta/E_{_J})^{1/2}$; the rate of
escape to this state is
\begin{equation}
\Gamma_{\textrm{out}}^{_{\ket{-}}}\sim
\frac{g_{_T}\delta_b}{4\pi}\sqrt{\frac{\Delta}{E_{_J}}}.
\label{gammaminus}
\end{equation}
These two rates are much higher than $\Gamma_{\textrm{in}}$
because $\delta_b\gg\delta_r$, so detours of the quasiparticle to
the CPB are short compared to the time quasiparticle spends in the
reservoir. Nevertheless, the typical time the quasiparticle spends
in the CPB is much greater than the oscillation period of the
qubit. Indeed, the ratio
\begin{equation}
\frac{\Gamma_{\textrm{out}}^{_{\ket{-}}}}{E_{_J}}\sim\frac{\delta_b}{\Delta}\sqrt{\frac{\Delta}{E_{_J}}}
\label{dephasing}
\end{equation}
is small: $\delta_b/\Delta\sim 10^{-4}-10^{-3}$ for any reasonable
size of the CPB (we used here the Ambegaokar-Baratoff relation
between $E_{_J}$, $g_{_T}$ and $\Delta$). The times of return of
the quasiparticle back to the reservoir are randomly distributed.
The probability of the quasiparticle returning to the reservoir
during times that are short compared to the oscillation period
$2\pi/E_{_J}$ is of the order $\Gamma_{\textrm{out}}/E_{_J}$ and
is small (here we do not distinguish between
$\Gamma_{\textrm{out}}^{_{\ket{-}}}$ and
$\Gamma_{\textrm{out}}^{_{\ket{+}}}$). Therefore, a single detour
of the quasiparticle into the CPB destroys coherent oscillations
of the qubit with overwhelming probability. Taking into account
the relation $\Gamma_{\textrm{in}}\ll\Gamma_{\textrm{out}}^\pm$,
we find that the dephasing rate for the qubit, induced by the
quasiparticle, is limited by the rate of quasiparticle tunneling
into the CPB
\begin{equation}
\frac{1}{T_2}\sim \Gamma_{\textrm{in}} \label{t2}
\end{equation}
with $\Gamma_{\textrm{in}}$ of Eq.~(\ref{gammain}).
 \begin{figure}
 \centering
 \includegraphics[width=3.0in]{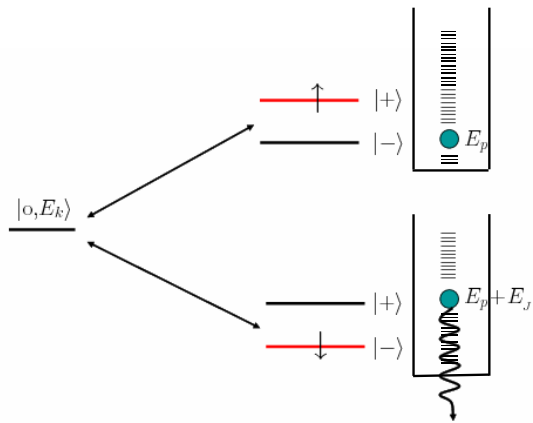}
 \caption{(color online). Schematic picture of the transitions
 between the qubit states in the presence of a quasiparticle in the reservoir, \emph{e.g.} $\ket{+,E_p}\leftrightarrow
 \ket{\mbox{\rm{o}},E_{k}}.$ Having kinetic energy $\sim E_{_J}$ the quasiparticle can emit a
 phonon. The corresponding state of the system is $\ket{-,E_p\!+\!E_{_J}}$.}
 \label{fig1}
 \end{figure}

 Unlike the phase, the energy stored in the degrees of freedom
 described by the qubit Hamiltonian (\ref{Hqubit}) is not dissipated
 at the short time scale given by Eq.~(\ref{t2}).  We start analyzing
 the time evolution of the qubit energy by considering the limit of
 infinitely slow quasiparticle relaxation (the latter typically is
 determined by the electron-phonon interaction~\cite{Kaplan,Clarke}). If
 initially the system was prepared in the $\ket{+}$ state, then upon a
 single cycle of quasiparticle tunneling, the qubit ends up in the
 ground state with a small probability defined by the ratio
 $\Gamma_{\textrm{out}}^{_{\ket{-}}}/\Gamma_{\textrm{out}}^{_{\ket{+}}}\sim\sqrt{T/E_{_J}}$.
 In other words, the qubit energy will randomly change over time
 between two values $\omega_+$ and $\omega_-$, see Eq.~(\ref{omega}).
 The portion of the time, the qubit spends in the ground state and the
 quasiparticle is excited to the energy $E_{_J}$ above the gap,
 is small $\sim\sqrt{T/E_{_J}}$.

 The portion of time that the quasiparticle spends in an excited state
 in the reservoir becomes important when we account for the
 phonon-induced relaxation of the quasiparticle. Having energy
 $E_{_J}$, the quasiparticle may emit a phonon at some rate $1/\tau$  and
 relax to a low-energy state. The relaxation of the
 quasiparticle will prevent further reexcitation of the qubit
 into $\ket{+}$ state and result in qubit energy relaxation. To find the energy
 relaxation rate of the qubit, we multiply the portion of time the
 quasiparticle spends in the excited state by the relaxation rate
 $1/\tau$:
\begin{equation}
\frac{1}{T_1}\sim\sqrt{\frac{T}{E_{_J}}}\frac{1}{\tau}. \label{t1}
\end{equation}
This estimate is applicable if $\tau\gg 1/\Gamma_{\textrm{in}}$,
and many cycles occur before the energy is dissipated into the
phonon bath.

 In the opposite case of fast relaxation $\tau\ll
1/\Gamma_{\textrm{in}}$, the quasiparticle loses its energy the
first time it gets it from the degrees of freedom of the
Hamiltonian (\ref{Hqubit}). Therefore, in this case the qubit
energy relaxation on average occurs on the time scale
\begin{equation}
\frac{1}{\widetilde{T}_1}\sim\sqrt{\frac{T}{E_{_J}}}\Gamma_{\textrm{in}}.
\label{tildet1}
\end{equation}

For aluminum, a typical superconductor used for charge qubits, the
quasiparticle relaxation time $\tau$ is indeed determined by the
inelastic electron-phonon scattering~\cite{Kaplan,Clarke}, and at
low energies ($\varepsilon \leq E_{_J}$) it can be estimated as
$\tau \! \gtrsim 10 \,\mu s$. For a small mesoscopic
superconductor this time is longer than the typical values of
$1/\Gamma_{\textrm{in}}$, and Eq.~(\ref{t1}) gives an adequate
estimate for the qubit energy relaxation rate.

A comparison of the phase relaxation time~(\ref{t2}) with even the
shortest of the two energy relaxation times~(\ref{tildet1}),
indicates that the coherence is destroyed much earlier than the
populations of the qubit states approach equilibrium. Therefore
one may consider the decay of qubit coherence separately from the
process of equilibration, which involves the electron-phonon
interaction in addition to the quasiparticle tunneling. In the
rest of the paper, we derive and solve the master equations, which
yield results discussed qualitatively in this section.

\section{Derivation of the Master Equations without quasiparticle
relaxation.}\label{derivation}

The Hamiltonian of the entire system consists of the qubit
Hamiltonian $H_{\textrm{qb}}$, BCS Hamiltonians for the
superconducting box and reservoir $H_{_{\textrm{BCS}}}^b$ and
$H_{_{\textrm{BCS}}}^r$, respectively, and quasiparticle tunneling
Hamiltonian $V$:
\begin{equation}
H=H_0+V,
\end{equation}
where
$H_0=H_{_{\textrm{BCS}}}^b+H_{_{\textrm{BCS}}}^r+H_{\textrm{qb}}$
and perturbation Hamiltonian $V$ takes into account only tunneling
of quasiparticles
 $V=H_{_T}\!-\!H_{_{_J}}$. The tunneling Hamiltonian $H_{_T}$ is
defined as
\begin{equation}\label{HT}
H_{_T}=\sum_{kp\sigma}(t_{kp}c_{k,\sigma}^{\dag}c_{p,\sigma}+\emph{h.c.}),
\end{equation}
where $t_{kp}$ is the tunneling matrix element, $c_{k,\sigma}$,
$c_{p,\sigma}$ are the annihilation operators for an electron in
the state $k,\sigma$ in the CPB and state $p,\sigma$ in the
superconducting reservoir, respectively; $H_{_J}$ is of the second
order in tunneling amplitude~\cite{footnote}
\begin{equation}
H_{_J}=\ket{N}\bra{N} H_{_T} \frac{1}{E-H_0}H_{_T}
\ket{N\!+2}\bra{N\!+2} +H.c.
\end{equation}
The matrix element $\bra{N} H_{_T} \frac{1}{E-H_0}H_{_T}
\ket{N+2}$ is proportional to effective Josephson energy $E_{_J}$,
and $H_{_T}$ is defined in Eq.~(\ref{HT}). Without quasiparticles
Hamiltonian $H_0$ reduces to Eq.~(\ref{Hqubit}), and qubit
dynamics can be described using the states $\ket{+}$ and
$\ket{-}$. In the presence of a quasiparticle, qubit phase space
should be extended. Relevant states now are $\ket{+,E_p}$,
$\ket{-,E_p}$, and $\ket{\mbox{\rm{o}},E_{k}}$. The first two
states $\ket{\pm,E_p}$ correspond to qubit being in the excited
(ground) state $\ket{\pm}$ and a quasiparticle residing in the
reservoir with energy $E_p=\sqrt{\xi_p^2+\Delta^2}$:
\[
\ket{\pm,E_p}\equiv\ket{\pm}\otimes\ket{E_p}.
\]
The third state $\ket{\mbox{\rm{o}},E_{k}}$ describes the qubit in
the ``odd'' state with $N\!+1$ electrons in the box,
\textit{i.e.}, the qubit escapes outside of its two-level Hilbert
space. Here $E_k=\sqrt{\xi_k^2+\Delta^2}$ is the energy of the
quasiparticle in the box. Perturbation Hamiltonian $V$ causes
transitions between the states $\ket{\pm,E_p}$ and
$\ket{\mbox{\rm{o}},E_{k}}$, see Fig.~\ref{fig1}. Note that $V$
does not induce the transitions between $\ket{+,E_p}$ and
$\ket{-,E_p}$.

The evolution of the full density matrix of the system is
described by Heisenberg equation of motion ($\hbar=1$):
\begin{eqnarray}\label{deni}
\dot{\rho_I}(t)=-i[V_{_I}(t),\rho_{_I}(t)],
\end{eqnarray}
where subscript $I$ stands for the interaction representation,
\emph{i.e.}, $V_{_I}(t)=e^{iH_0t}Ve^{-iH_0t}$. The iterative
solution of Eq.~(\ref{deni}) yields for the matrix elements of the
density matrix
\begin{eqnarray}\label{master2}
\!\bra{s}\dot{\rho}_{_I}(t)\!\ket{s'}\!&\!=
\!&\!-i\bra{s}[V_{_I},\rho(0)]\ket{s'}\\
\!&\!-\!&\!\int_{0}^{t}\!{d\tau}\!\bra{s}\!
\left[V_{_I}(t),[V_{_I}(t\!-\!\tau),\rho_{_I}(t\!-\!\tau)]\right]\!\ket{s'}\!,\nonumber
\end{eqnarray}
where $\ket{s}$ can be $\ket{+,E_p}$, $\ket{-,E_{p}}$, or
$\ket{\mbox{\rm{o}},E_{k}}$. The interaction Hamiltonian $V$ has
no diagonal elements in the representation for which $H_0$ and
$\rho(0)$ are diagonal. Therefore, the first term in right-hand
side of Eq.~(\ref{master2}) is equal to zero,
\begin{eqnarray}\label{master1}
\!\bra{s}\dot{\rho}_{_I}(t)\!\ket{s'}\!\!=
\!-\!\!\int_{0}^{t}\!{d\tau}\!\bra{s}\!
\left[V_{_I}(t),[V_{_I}(t\!-\!\tau),\rho_{_I}(t\!-\!\tau)]\right]\!\ket{s'}\!.\nonumber\\
\end{eqnarray}
Equation (\ref{master1}) implies that evolution of the projected
density matrix is proportional to $V^2$. Since interaction is
assumed to be weak, the rate of change of $\rho_{_I}(t\!-\!\tau)$
is slow compared to that of $V_{_I}(t)$. Therefore, one can
approximate $\rho_{_I}(t\!-\!\tau)$ by $\rho_{_I}(t)$ in the
right-hand side of Eq.~(\ref{master1}) (see, for example,
Refs.~[\onlinecite{Zubarev, Esposito}] for more details on the
derivation). Finally, going back to the original representation,
we arrive at the following system of master equations
\begin{eqnarray}\label{master4}
\!&\!\bra{s}\!&\!\dot{\rho}(t)\ket{s'}=-\!i\bra{s}(E_s\!-\!E_{s'})\rho(t)\ket{s'}\nonumber\\
\!&\!-\!&\!\pi\sum_{m,n}\bra{s}V\ket{m}\bra{m}V\ket{n}\bra{n}\rho(t)\ket{s'}\delta(E_n\!-\!E_m)\nonumber\\
\!&\!-\!&\!\pi\sum_{m,n}\bra{s}\rho(t)\ket{m}\bra{m}V\ket{n}\bra{n}V\ket{s'}\delta(E_{m}\!-\!E_n)\nonumber\\
\!&\!+\!&\!\pi\sum_{m,n}\bra{s}V\ket{m}\bra{m}\rho(t)\ket{n}\bra{n}V\ket{s'}\delta(E_n\!-\!E_{s'})\nonumber\\
\!&\!+\!&\!\pi\sum_{m,n}\bra{s}V\ket{m}\bra{m}\rho(t)\ket{n}\bra{n}V\ket{s'}\delta(E_m\!-\!E_s),\nonumber\\
\end{eqnarray}
where states $\ket{m}$, $\ket{n}$, $\ket{s}$, and $\ket{s'}$
denote $\ket{+,E_p}$, $\ket{-,E_{p}}$, or
$\ket{\mbox{\rm{o}},E_{k}}$, and the sum runs over all possible
configurations. The system of equations~(\ref{master4}) describes
the kinetics of the qubit in the presence of a quasiparticle in
the Markovian approximation. We are interested in elements of the
density matrix that are diagonal in quasiparticle subspace,
\textit{e.g.},
$P_{_{\!+-\!}}(E_p,t)=\bra{\!+,E_p}\rho(t)\ket{E_p,-}$, since at
the end one should take the trace over quasiparticle degrees of
freedom to obtain observable quantities. Note that
Eq.~(\ref{master4}) which describes evolution of a closed system
(the qubit and the quasiparticle) does conserve its total energy.
We will include the mechanisms of energy loss to the phonon bath
and discuss the proper modifications of the master equation later
in Sec.~\ref{withrelax}.

We now apply secular approximation to Eq.~(\ref{master4}). This is
justified due to the separation of the characteristic time scales
$E_{_J}^{-1}\ll \Gamma_{\textrm{out}}^{-1} \ll
\Gamma_{\textrm{in}}^{-1}$ established in the previous section.
When considering the evolution of the off-diagonal elements of the
density matrix $P_{_{\!+-\!}}(E_p,t)$, we need to keep only terms
$\propto P_{_{\!+-\!}}(E_p,t)$ in the right-hand-side of the
corresponding master equation. The contribution of other elements
of the density matrix to the evolution of the coherences
$P_{_{\!+-\!}}(E_p,t)$ is small as $\Gamma_{\textrm{out}}/E_{_J}$
and $\Gamma_{\textrm{in}}/E_{_J}$. Thus, we arrive at the equation
governing the evolution of the coherences
\begin{eqnarray}\label{coherences}
\dot{P}_{_{{\!+-\!}}}(E_p,t)\!&\!=\!&\!-iE_{_J}P_{_{\!+-\!}}(E_p,t)\nonumber\\
\!&\!-\!&\!\frac{1}{2}\sum_k \left[W_{\textrm{+}}(E_{p},E_k)+
W_{\textrm{-}}(E_{p},E_k)\right]P_{_{\!+-\!}}(E_p,t).\nonumber\\
\end{eqnarray}
The transition rates $W_{\textrm{+}}(E_{p},E_k)$ and
$W_{\textrm{-}}(E_{p},E_k)$ are given by the Fermi golden rule
\begin{eqnarray}\label{rates}
W_{\textrm{+}}(E_p,\!E_k)\!=\!2\pi|\bra{E_p,\!+}H_{_T}\ket{\mbox{\rm{o}},E_k}|^2\delta(E_p\!+\!\omega_{+}\!-\!E_k) \nonumber\\
W_{\textrm{-}}(E_p,\!E_k)\!=\!2\pi|\bra{E_p,\!-}H_{_T}\ket{\mbox{\rm{o}},E_k}|^2\delta(E_p\!+\!\omega_{-}\!-\!E_k)
\nonumber\\
\end{eqnarray}
with $\omega_{\pm}$ of Eq.~(\ref{omega}). The matrix elements
$\bra{E_p,\pm}H_{T}\ket{\mbox{\rm{o}},E_k}$ can be calculated
using the particle conserving version of the Bogoliubov
transformation~\cite{Schrieffer} and are
$\bra{E_p,\pm}H_{T}\ket{\mbox{\rm{o}},E_k}=\sqrt{2}(t_{pk}u_p u_k
- t_{kp}v_p v_k)$, where $u_p$, $v_p$ are Bogoliubov coherence
factors:
\begin{eqnarray}
u_p^2=\frac{1}{2}\left(1+\frac{\xi_{p}}{E_p}\right),
\mbox{ and }
v_p^2=\frac{1}{2}\left(1-\frac{\xi_{p}}{E_p}\right),
\end{eqnarray}
leading to~\cite{Tinkham}
\begin{eqnarray}\label{rates2}
W_{\textrm{+}}(E_p,\!E_k)\!=\!2\pi|t_{pk}|^2\!\left(\!1\!+\!\frac{\xi_p\xi_k\!-\!\Delta^2}{E_pE_k}\!\right)\!\delta(E_p\!+\!\omega_{+}\!-\!E_k), \nonumber\\
W_{\textrm{-}}(E_p,\!E_k)\!=\!2\pi|t_{pk}|^2\!\left(\!1\!+\!\frac{\xi_p\xi_k\!-\!\Delta^2}{E_pE_k}\!\right)\!\delta(E_p\!+\!\omega_{-}\!-\!E_k).
\nonumber\\
\end{eqnarray}
Now we may relate the tunneling matrix elements to the
normal-state junction conductance
\begin{eqnarray}
g_{_T} =8\pi^2 \sum_{p,k}|t_{pk}|^2 \delta(\xi_p)\delta(\xi_k)
\nonumber.
\end{eqnarray}
Assuming that tunnel matrix elements $t_{pk}$ are weakly dependent
on the energies $\xi_k$, $\xi_p$, we can rewrite Eq.~(\ref{rates2}) in terms of the
dimensionless conductance:
\begin{eqnarray}\label{rates3}
W_{\textrm{+}}(E_p,\!E_k)\!=\!\frac{g_{_T}\delta_r \delta_b}{4\pi}\!\left(\!1\!+\!\frac{\xi_p\xi_k\!-\!\Delta^2}{E_pE_k}\!\right)\!\delta(E_p\!+\!\omega_{+}\!-\!E_k), \nonumber\\
W_{\textrm{-}}(E_p,\!E_k)\!=\!\frac{g_{_T}\delta_r
\delta_b}{4\pi}\left(\!1\!+\!\frac{\xi_p\xi_k\!-\!\Delta^2}{E_pE_k}\!\right)\!\delta(E_p\!+\!\omega_{-}\!-\!E_k),
\nonumber\\
\end{eqnarray}
where $\delta_{r(b)}$ is mean level spacing in the reservoir(box)
$\delta_{r(b)}=(\nu_{r(b)} V_{r(b)})^{-1}$ with $\nu_{r(b)}$ being
single-spin electron density of states at the Fermi level in the
reservoir~(box).

The system of equations for the diagonal part of the density matrix
describes the evolution of the populations and follows from
Eq.~(\ref{master4}). From now on we adopt the short-hand notation for
the diagonal elements of the density matrix
$\bra{s}\rho(t)\ket{s}=P_{ss}(E_s,t)$. In particular, we denote the
probability of the qubit to be in the state $\ket{+}$ or in the state
$\ket{-}$ and a quasiparticle to have energy $E_p$ as $P_{++}(E_p,t)$
or $P_{--}(E_p,t)$, respectively; the probability
$P_{\textrm{o}}(E_k,t)$ corresponds to the state with a quasiparticle
residing in the CPB and having energy $E_k$. In these notations the
system of equations describing the dynamics of the populations for the
states $\ket{+,E_p}$, $\ket{-,E_{p}}$ or $\ket{\mbox{\rm{o}},E_{k}}$
can be written as
\begin{subequations}\label{mastermain}
\begin{equation}\label{mastermain_a}
\!\dot{P}_{_\textrm{++}}\!(\!E_p,\!t)\!\!+\!\sum_{k}W_{\textrm{+}}(E_p,E_k)\left[P_{_\textrm{++}}(E_p,t)\!-\!P_{\textrm{o}}(E_k,t)\right]\!=\!0,\!\\
\end{equation}
\begin{equation}\label{mastermain_b}
\!\dot{P}_{\textrm{-\,-}}(E_p,\!t)\!+\!\sum_{k}W_{\textrm{-}}(E_p,E_k)\left[P_{\textrm{-\,-}}(E_p,t)\!-\!P_{\textrm{o}}(E_k,t)\right]\!=0,\!\\
\end{equation}
\begin{eqnarray}\label{mastermain_c}
\!\dot{P}_{\textrm{o}}(E_k,\!t)\!\!+\!\!\sum_{p}\left[W_{\textrm{\!+\!}}(E_p,E_k)\!+\!W_{\textrm{-}}(E_p,E_k)\right]P_{\textrm{o}}(E_k,t)\!-\!\nonumber\\
\!-\!\sum_{p}\!\left[W_{\textrm{+}}\!(\!E_p,\!E_k)P_{_\textrm{\!++\!}}\!(\!E_p,\!t)\!+\!W_{\textrm{-}}\!(\!E_p,\!E_k)P_{\textrm{-\,-}}\!(\!E_p,\!t)\right]\!=\!0,\!\nonumber\\
\end{eqnarray}
\end{subequations}
where we neglected the contribution of the coherences. This is
justified as long as $P_{_{\!+-\!}}(E_p,0)=0$ in the initial
moment of time, and the two parameters
$\Gamma_{\textrm{out}}/E_{_J}$ and $\Gamma_{\textrm{in}}/E_{_J}$,
are small. The transition rates in Eqs.~(\ref{mastermain}) are
given by the Fermi golden rule, see Eqs.~(\ref{rates}).

At the end, experimentally observable quantities can be obtained
from $P_{ij}$ by taking the proper trace over the quasiparticle
degrees of freedom
\begin{eqnarray}\label{def_sigma}
\sigma_{_{ij}}(t)=\sum_pP_{_{ij}}(E_p,t),
\end{eqnarray}
where $i,j=+,-$. This completes the derivation of the master
equations without quasiparticle relaxation, and we proceed to the
solution of these equations.

\section{Evolution of the qubit coherences.} \label{seccoher}
We now discuss the solution for the off-diagonal elements of the
density matrix. We assume that initially the qubit and
quasiparticle are independent; the quasiparticle is in thermal
equilibrium in the reservoir, and the qubit is prepared in a
superposition state with $\sigma_{_{\!+-\!}}(0)\neq 0$:
\begin{eqnarray}\label{IC_coh}
P_{_{\!+-\!}}(E_p,0)=\rho_{\rm{odd}}(E_p)\sigma_{_{\!+-\!}}(0)\neq 0,
\end{eqnarray}
where $\rho_{\rm{odd}}(E_p)$ is the equilibrium distribution
function with an odd number of electrons in reservoir at
temperature $T\ll \Delta$,
\begin{eqnarray}
\rho_{\rm{odd}}(E_p)=\frac{\exp\left(-E_p/T\right)}{Z_{\rm{odd}}}.
\end{eqnarray}
The normalization factor $Z_{\rm{odd}}$ here is
$$Z_{\rm{odd}}=\sum_p\exp\left(-\frac{E_p}{T}\right)=\sqrt{\frac{\pi}{2}}\frac{\Delta}{\delta_r}\sqrt{\frac{T}{\Delta}}\exp\left(-\frac{\Delta}{T}\right).$$

The solution of Eq.~(\ref{coherences}) is straightforward. After
tracing out quasiparticle degrees of freedom we obtain
\begin{eqnarray}\label{coherence}
\!\sigma_{_{\!+-\!}}\!(t)\!=\!\sigma_{_{\!+-\!}}\!(0)\!\sum_p\rho_{\rm{odd}}(E_p)\!\exp\!\left(\!-iE_{_J}t\!-\!\frac{1}{2}\Gamma_{\textrm{in}}(E_p)t\!\right)\!,
\end{eqnarray}
where $\Gamma_{\textrm{in}}(E_p)$ is given by
\begin{eqnarray}
\Gamma_{\textrm{in}}(E_p)=\sum_k \left[W_{\textrm{+}}(E_{p},E_k)+
W_{\textrm{-}}(E_{p},E_k)\right].
\end{eqnarray}
In the low-temperature limit $T \ll \omega_{-},\omega_{+}$, the
expression for $\sigma_{_{\!+-\!}}(t)$ can be simplified
\begin{eqnarray}
\!\sigma_{_{\!+-\!}}(t)\!=\!\sigma_{_{\!+-\!}}(0)\!\exp\!\left(-iE_{_J}t\!-\!\frac{t}{T_2}\right).
\end{eqnarray}
Here the phase relaxation time $T_2$ is given by
\begin{eqnarray}\label{T22}
T_2=\frac{g_{_T}\delta_r}{8\pi}\left(\sqrt{\frac{\omega_{_+}}{2\Delta+\omega_{_+}}}+\sqrt{\frac{\omega_{_-}}{2\Delta+\omega_{_-}}}\,\right).
\end{eqnarray}

The decay of qubit coherences is determined by the rate of the
quasiparticle tunneling into the box as previously discussed in
Sec.~\ref{quali}. This result remains valid also in the presence
of quasiparticle relaxation. On the contrary, the evolution of the
diagonal parts of the density matrix $P_{++}$ and $P_{--}$,
depends strongly on the relaxation of the quasiparticles. We will
study this evolution with and without quasiparticle relaxation in
the next sections.

\section{Kinetics of the qubit populations without quasiparticle
relaxation}\label{norel}
 The evolution of the diagonal elements
of the density matrix is described by Eq.~(\ref{mastermain}). We
will assume that initially the qubit is prepared in the state
$\ket{+}$, and the quasiparticle resides in the reservoir. As
explained in Sec.~\ref{quali} tunneling out of the box ($\sim\!\!
\Gamma_{\textrm{out}}$) is much faster than tunneling in ($\sim
\!\!\Gamma_{\textrm{in}}$) due to the differences in the volumes
of the CPB and the reservoir. In fact, for a sufficiently small
box, $1/\Gamma_{\textrm{out}}$ is the shortest time scale in the
system of Eqs.~(\ref{mastermain}). Therefore, we may neglect term
$\partial_t P_{\textrm{o}}(E_k,t)$ in Eq.~(\ref{mastermain_c});
\emph{i.e.}, the value of $P_{\textrm{o}}(E_k,t)$ follows
instantaneously the time variations of $P_{_{++}}(E_p,t)$ and
$P_{_{--}}(E_p,t)$. This greatly simplifies the system of
equations for the populations. The solution for
$P_{\textrm{o}}(E_k,t)$ in this approximation is
\begin{eqnarray}\label{podd}
\!P_{\textrm{o}}(E_k,t)&=&\frac{\sum_{p}W_{_\textrm{+}}(E_p,E_k)
P_{_\textrm{++}}(E_p,t)}{\sum_{p}\left[W_{_\textrm{\!+\!}}(E_p,E_k)\!+\!W_{_\textrm{-}}(E_p,E_k)\right]}\nonumber\\
&+&\frac{\sum_{p}W_{_\textrm{-}}(E_p,E_k)P_{_\textrm{-\,-}}(E_p,t)}{\sum_{p}\left[W_{_\textrm{\!+\!}}(E_p,E_k)\!+\!W_{_\textrm{-}}(E_p,E_k)\right]}.
\end{eqnarray}
After substituting this expression back into
Eqs.~(\ref{mastermain_a}) and ~(\ref{mastermain_b}), we obtain
effective rate equations for the qubit in the presence of an
unpaired electron in the superconducting parts
\begin{eqnarray}\label{kin}
\!\dot{P}_{_\textrm{++}}(E_p,t)\!&\!+\!&\!\gamma_{_+}(E_p)P_{_\textrm{++}}(E_p,\!t)\!-\!\gamma_{_+}(E_p)P_{\textrm{-\,-}}(E_p\!+\!E_{_J},\!t)\!=\! 0, \nonumber\\\nonumber\\
 \nonumber
 \!\dot{P}_{\textrm{-\,-}}(E_{p},t)\!&\!+\!&\!\gamma_{_-}(E_{p})P_{\textrm{-\,-}}(E_{p},\!t)\!-\!\gamma_{_-}(E_{p})P_{_\textrm{++}}(E_{p}\!-\!E_{_J},\!t)\!=\! 0 \\
\end{eqnarray}
with $\gamma_{_\pm}(E_p)$ having the form
\begin{eqnarray}\label{ratesnew}
\gamma_{_+}(E_p)\!&\!=\!&\!\sum_{p',k}\frac{W_{\textrm{+}}(E_p,E_k)W_{\textrm{-}}(E_{p'},E_k)}{\sum_{p''}[W_{\textrm{+}}(E_{p''},E_k)\!+\!W_{\textrm{-}}(E_{p''},E_k)]}\nonumber,\\
\gamma_{_-}(E_{p})\!&\!=\!&\!\sum_{p',k}\frac{W_{\textrm{-}}(E_p,E_k)W_{\textrm{+}}(E_{p'},E_k)}{\sum_{p''}[W_{\textrm{+}}(E_{p''},E_k)\!+\!W_{\textrm{-}}(E_{p''},E_k)]}.\nonumber\\
\end{eqnarray}
This structure of the transition rates reflects the nature of the
transitions involving an intermediate state
$\ket{\mbox{\rm{o}},E_{k}}$. The normalization condition
\begin{eqnarray}\label{norm}
\sum_p [P_{_\textrm{++}}(E_p,\!t)+P_{\textrm{-\,-}}(E_p,\!t)]=1
\end{eqnarray}
 is preserved under evolution. This can be checked directly with the help of
Eqs.~(\ref{kin}) and the following relation for the rates:
\begin{eqnarray}\label{balance}
\sum_p\gamma_{_+}(E_{p})X(E_p)=\sum_p\gamma_{_-}(E_{p})X(E_p-E_{_J})
\end{eqnarray}
(here $X(E_p)$ is an arbitrary function of $E_p$).

Let us discuss the solution of the Eqs.~(\ref{kin}). In the
initial moment of time the qubit and quasiparticle are
uncorrelated; the qubit is prepared in the excited state $\ket{+}$
and quasiparticle can be described by the equilibrium distribution
function $\rho_{\rm{odd}}(E_p)$:
\begin{eqnarray}\label{IC}
P_{_\textrm{++}}(E_p,0)\!=\!\rho_{\rm{odd}}(E_p) \ \mbox{ \ and \
} \ P_{_\textrm{-\,-}}(E_{p}\!+\!E_{_J},0)\!=\!0.
\end{eqnarray}
Upon solving Eqs.~(\ref{kin}), we find expressions for the
populations of the qubit levels
\begin{eqnarray}\label{sigma35}
\sigma_{_{++}}(t)\!&\!=\!&\!\sum_{p}\!\frac{\rho_{\rm{odd}}(E_p)\gamma_{_-}(E_{p}\!+\!E_{_J})}{\gamma_{_-}(E_{p}\!+\!E_{_J})+\gamma_{_+}(E_p)}\!+\!\nonumber\\
\!&\!+\!&\!\sum_{p}\frac{\rho_{\rm{odd}}(E_p)\gamma_{_+}(E_p)}{\gamma_{_-}(E_{p}\!+\!E_{_J})+\gamma_{_+}(E_p)}\exp\left(-\Gamma(E_p)t\right),\nonumber\\
\nonumber\\
\sigma_{_{-\,-}}(t)&\!=\!&\sum_{p}\frac{\rho_{\rm{odd}}(E_p\!-\!E_{_J})\gamma_{_-}(E_{p})}{\gamma_{_-}(E_{p})+\gamma_{_+}(E_p-E_{_J})}\!-\!\nonumber\\
\!&\!-\!&\!\sum_{p}\frac{\rho_{\rm{odd}}(E_p\!-\!E_{_J})\gamma_{_-}(E_{p})}{\gamma_{_-}(E_{p})+\gamma_{_+}(E_p-E_{_J})}\exp\left(-\Gamma(E_p\!-\!E_{_J})t\right)\nonumber,\\
\end{eqnarray}
where $\Gamma(E_p)$ is defined as
\begin{eqnarray}\label{Gamma}
\Gamma(E_p)=\gamma_{_+}(E_p)+\gamma_{_-}(E_{p}\!+\!E_{_J}).
\end{eqnarray}
In the low-temperature limit, we calculate the sums in
Eq.~(\ref{sigma35}) assuming $\Delta>\omega_{_+}> E_{_J}\gg T$ to
find
\begin{eqnarray}\label{leading}
\sum_p \frac{\gamma_{_+}(E_p) \rho_{\rm{odd}}(E_p)}{\Gamma(E_p)}
\approx \sqrt{\frac{T}{\pi E_{_J}}}
\end{eqnarray}
and
\begin{eqnarray}
\sigma_{_{++}}(t)\!&\!=\!&\!1-\sqrt{\frac{T}{\pi E_{_J}}}+\sqrt{\frac{T}{\pi E_{_J}}}\exp\left[-\frac{t}{T_1^{*}}\right],\nonumber\\
\nonumber\\
\sigma_{_{-\,-}}(t)&\!=\!&\sqrt{\frac{T}{\pi E_{_J}}}\left(1-\exp\left[-\frac{t}{T_1^{*}}\right]\right).\nonumber\\
\end{eqnarray}
The relaxation time $T_1^{*}$ is defined as
\begin{eqnarray}
\frac{1}{T_1^{*}}\approx\frac{g_{_T}\delta_r}{4\pi}\sqrt{\frac{\omega_{+}}{2\Delta+\omega_{+}}}\left(1+\frac{E_{_J}}{\omega_{+}}\right).
\end{eqnarray}
In deriving this expression we assumed $E_{_J}/\Delta\ll 1$ and
kept only the leading terms.

The solution for the populations in this case (no quasiparticle
relaxation, $\tau=\infty$) show that final qubit populations are
determined by the tunneling rates, which, in turn, depend on the
superconducting DOS at different energies $\nu(E_p)$ and
$\nu(E_p\!+\!E_{_J})$. Since the states with higher DOS are more
favorable, the quasiparticle can be found most of the time with
energy close to $\Delta$ and rarely with energy $\Delta+E_{_J}$.
Therefore, at low temperatures $T \ll E_{_J}$, the qubit will
mostly remain in the excited state $\ket{+}$. The probability to
find the qubit in the ground state is proportional to $\sim
\sqrt{T/E_{_J}}$ and thus is small, see Sec.~\ref{quali}. As soon
as we include mechanisms of quasiparticle relaxation into
consideration, the qubit populations will eventually reach
equilibrium. In the next sections we investigate the equilibration
of the qubit.

\section{ Kinetics of the Qubit populations with quasiparticle
relaxation in the reservoir}\label{withrelax}

\subsection{Master equations with quasiparticle
relaxation}\label{withrelax_a}

Here we consider a more realistic model by incorporating the
mechanisms of quasiparticle relaxation into the rate equations.
Such mechanisms were studied in the context of non-equilibrium
superconductivity~\cite{Kaplan,Clarke}. In aluminum, a typical
superconductor used in charge qubits, the dominant mechanism of
quasiparticle relaxation is due to inelastic electron-phonon
scattering. The relaxation time depends on the excess energy
$\varepsilon$ of a quasiparticle~\cite{Kaplan}
\begin{eqnarray}
\frac{1}{\tau(\varepsilon)}=
\frac{1}{\tau_0}\frac{64\sqrt{2}}{105}\left(\frac{\Delta}{T_c}\right)^3\left(\frac{\varepsilon}{\Delta}\right)^{7/2},
\end{eqnarray}
where $\varepsilon=E_p-\Delta\ll\Delta$, and $\tau_0$ is
characteristic parameter defining electron-phonon scattering rate
at $T=T_c$ (here $T_c$ is superconducting transition temperature).
For typical excess energies of the order of $E_{_J}\sim 0.3$ K,
the estimate for $\tau$ yields  quite long relaxation time $\tau
\! \sim\! 10^{-5}\!-\!10^{-4}$~s.

The procedure developed in Sec.~\ref{derivation} allows us to
include the mechanisms of quasiparticle relaxation into the master
equations. One can start by writing an equation of motion for the
density matrix that includes the qubit, quasiparticle, and
phonons, then expand the density matrix in the small coupling
parameter - electron-phonon interaction as discussed in
Sec.~\ref{derivation}. Finally, one should trace out phonon
degrees of freedom and obtain master equations for the qubit with
quasiparticle relaxation. We will skip the cumbersome derivation
and present only the results here. In the relaxation time
approximation the collision integral has the form
\begin{eqnarray}
\mathcal{I_{\pm}}=-\frac{1}{\tau}\left(P_{_{\pm\pm}}(E_p,t)\!-\!\overline{P}_{_{\pm\pm}}(E_p,t)\right),
\label{collint}
\end{eqnarray}
where $\tau=\!\tau(\varepsilon\!\sim\!E_{_J}\!)$. The probability
$\overline{P}_{_{\pm\pm}}(E_p,t)$ is proportional to the
equilibrium distribution function of a quasiparticle
$\rho_{\rm{odd}}(E_p)$ and the proper qubit population
$\sigma_{_{\pm\pm}}(t)$:
\begin{eqnarray}\label{equilP}
\overline{P}_{_{\pm\pm}}(E_p,t)=\rho_{\rm{odd}}(E_p)\sum_p{P_{_{\pm\pm}}(E_p,t)}.
\end{eqnarray}
The form of $\overline{P}_{_{\pm\pm}}(E_p,t)$ is dictated by the
fact that phonons equilibrate the quasiparticle only, without
affecting directly the qubit states~\cite{phonons,Ioffe}. The
collision integral Eq.~(\ref{collint}) replaces zero in the
right-hand sides of Eqs.~(\ref{mastermain_a})
and~(\ref{mastermain_b}). However, Eq.~(\ref{mastermain_c}) for
$P_{\textrm{o}}(E_k,t)$ remains unchanged due to the short
dwelling time of a quasiparticle in the box (we assume that
$\tau(\varepsilon\!\sim\!\omega_{+}\!) \gg
\Gamma_{\textrm{out}}^{-1}$, but set no constraints on $\tau
\Gamma_{\textrm{in}}$). Then, the system of Eqs.~(\ref{kin}) for
populations can be written as
\begin{eqnarray}\label{kin_nq}
 \dot{P}_{_\textrm{++}}(E_p,t)\!&\!+\!&\!\gamma_{_+}(E_p)P_{_\textrm{++}}(E_p,t)\!-\!\gamma_{_+}(E_p)P_{\textrm{-\,-}}(E_p\!+\!E_{_J},t)\!=\nonumber\\
 \!&\!-\!&\!\!\frac{1}{\tau}\left(P_{_\textrm{++}}(E_p,t)\!-\!\overline{P}_{_\textrm{++}}(E_p,t)\right),\nonumber\\
 \nonumber
\\
 \dot{P}_{\textrm{-\,-}}(E_{p},t)\!&\!+\!&\!\gamma_{_-}(E_{p})P_{\textrm{-\,-}}(E_{p},t)\!-\!\gamma_{_-}(E_{p})P_{_\textrm{++}}(E_{p}\!\!-\!E_{_J},t)\!=\nonumber\\
 \!&\!-\!&\!\frac{1}{\tau}\left(P_{\textrm{-\,-}}(E_{p},t)\!\!-\!\!\overline{P}_{_\textrm{-\,-}}(E_{p},t)\right)\\
\nonumber
 \end{eqnarray}
with the effective transition rates $\gamma_{_\pm}(E_p)$ defined
in Eq.~(\ref{ratesnew}). The obtained system of
integro-differential equations~(\ref{kin_nq}) for
$P_{_{\pm\pm}}(E_p,t)$ describes the effect of quasiparticle
relaxation on the dynamics of the qubit.

We solve Eqs.~(\ref{kin_nq}) first in the simple case of a short
relaxation time ($\tau \ll \Gamma_{\textrm{in}}^{-1}$).  Under
these assumptions, we can seek the solution in the form
\begin{equation}\label{anzats}
P_{_{\pm\pm}}(E_p,t)=\rho_{\rm{odd}}(E_p)\ \sigma_{_{\pm\pm}}(t),
\end{equation}
with $\sigma_{_{\pm\pm}}(t)$ defined in Eq.~(\ref{def_sigma}), so
that $P_{_{\pm\pm}}(E_p,t)=\overline{P}_{_{\pm\pm}}(E_p,t)$. Using
this ansatz and performing the appropriate summation,
  Eqs.~(\ref{kin_nq}) reduce to the Bloch-Redfield equations
\begin{eqnarray}\label{Bloch-Redf}
 \dot{\sigma}_{_\textrm{++}}(t)&+&\sum_{p}\gamma_{_+}(E_p)\rho_{\rm{odd}}(E_p)\sigma_{_\textrm{++}}(t)-\nonumber\\
 &-&\sum_{p}\gamma_{_+}(E_p)\rho_{\rm{odd}}(E_p+E_{_J})\sigma_{\textrm{-\,-}}(t)=0,\nonumber\\
 \dot{\sigma}_{_\textrm{-\,-}}(t)&+&\sum_{p}\gamma_{_-}(E_p)\rho_{\rm{odd}}(E_p)\sigma_{\textrm{-\,-}}(t)-\nonumber\\
 &-&\sum_{p}\gamma_{_-}(E_p)\rho_{\rm{odd}}(E_p-E_{_J})\sigma_{_\textrm{++}}(t)=0.\nonumber\\
\end{eqnarray}
Utilizing the property of the rates Eq.~(\ref{balance}), one can
simplify the equations above,
\begin{eqnarray}\label{fast}
 \dot{\sigma}_{_\textrm{++}}(t)&+&\left<\!\gamma_{_+}\!\right>\sigma_{_\textrm{++}}(t)=\left<\!\gamma_{_-}\!\right>\sigma_{\textrm{-\,-}}(t),\nonumber\\ \nonumber\\
 \dot{\sigma}_{\textrm{-\,-}}(t)&+&\left<\!\gamma_{_-}\!\right>\sigma_{\textrm{-\,-}}(t)=\left<\!\gamma_{_+}\!\right>\sigma_{_\textrm{++}}(t),
\end{eqnarray}
where thermal-averaged transition rates
$\left<\!\gamma_{_+}\!\right>$ and $\left<\!\gamma_{_-}\!\right>$
are
\begin{eqnarray}\label{rates_fast}
\left<\!\gamma_{_+}\!\right>=\sum_{p}\gamma_{_+}(E_p)\rho_{\rm{odd}}(E_p),\nonumber\\
\left<\!\gamma_{_-}\!\right>=\sum_{p}\gamma_{_-}(E_p)\rho_{\rm{odd}}(E_p).
\end{eqnarray}
One can also check that due to relation~(\ref{balance}) rates
$\left<\gamma_{_\pm}\right>$ comply with the detailed balance
requirement
$$\frac{\left<\!\gamma_{_+}\!\right>}{\left<\!\gamma_{_-}\!\right>}=\exp\left(\frac{E_{_J}}{T}\right).$$

For the initial conditions: $\sigma_{_\textrm{++}}(0)=1$,
$\sigma_{\textrm{-\,-}}(0)=0$, the solution for populations is
\begin{eqnarray}\label{sol_sigma_tau0}
\sigma_{_\textrm{++}}(t) \!&\!=\!&\!\frac{e^{- E_{_J}/T}}{1+e^{-
E_{_J}/T}}+\frac{e^{-\left(\left<\!\gamma_{_+}\!\right>+\left<\!\gamma_{_-}\!\right>\right)t}}{1+e^{-
E_{_J}/T}},
\nonumber\\\nonumber\\
\sigma_{\textrm{-\,-}}(t) \!&\!=\!&\!\frac{1}{1+e^{-
E_{_J}/T}}\left[1-e^{-\left(\left<\!\gamma_{_+}\!\right>+\left<\!\gamma_{_-}\!\right>
\right)t}\right].
\end{eqnarray}
In the low-temperature limit ($T\ll E_{_J}<\omega_{_+}<\Delta$) we
find a simple form for the effective rates
$\left<\!\gamma_{_\pm}\!\right>$ in the leading order in
$T/\Delta$ and $E_{_J}/\Delta$:
\begin{eqnarray}\label{rates_fast2}
\left<\!\gamma_{_-}\!\right>\!&\!=\!&\!\frac{g\delta_r}{4\pi}\sqrt{\frac{\omega_{_+}}{2\Delta+\omega_{_+}}}\left(1+\frac{E_{_J}}{\omega_{_+}}\right)\sqrt{\frac{T}{\pi
E_{_J}}}\exp\left(-\frac{E_{_J}}{T}\right)\nonumber,\\
\left<\!\gamma_{_+}\!\right>\!&\!=\!&\!\frac{g\delta_r}{4\pi}\sqrt{\frac{\omega_{_+}}{2\Delta+\omega_{_+}}}\left(1+\frac{E_{_J}}{\omega_{_+}}\right)\sqrt{\frac{T}{\pi
E_{_J}}}.
\end{eqnarray}
Factors of $\sqrt{T/\pi E_{_J}}$ in the rates can be interpreted
as the probability of flipping the qubit
($\ket{+}\rightarrow\ket{-}$), which is mainly determined by the
ratio of the DOS of quasiparticles at energies $\nu(E_p)$ and
$\nu(E_p\!+\!E_{_J})$, respectively (see the discussion in
Sec.~\ref{quali}). We would like to point out here that energy
relaxation rate
$\left<\!\gamma_{_+}\!\right>+\left<\!\gamma_{_-}\!\right>$ is
smaller than the phase relaxation rate by a factor of $\sqrt{T/\pi
E_{_J}}$.

\subsection{ General solution for the qubit populations in the
relaxation time approximation} In this section we find the
solution of Eqs.~(\ref{kin_nq}) at an arbitrary value of
$\Gamma_{\textrm{in}}\tau$. In order to find the solution for the
qubit populations we will use Laplace transform
\begin{eqnarray}
P(E_p,s)=\int_0^{\infty}{dt}P(E_p,t)e^{-st},
\end{eqnarray}
and reduce the system of differential equations~(\ref{kin_nq})
supplied with the initial conditions Eq.~(\ref{IC}) to the system
of algebraic equations
\begin{eqnarray}\label{algeb}
 \!sP_{_\textrm{++}}\!(\!E_p,\!s)\!&\!-\!&\!\rho_{\rm{odd}}(\!E_p)\!+\!\gamma_{_+}(\!E_p)P_{_\textrm{++}}\!(\!E_p,\!s)\!-\!\gamma_{_+}(\!E_p)\widetilde{P}_{\textrm{-\,-}}(E_p,\!s)\nonumber\\
 \! & \!=\! & \!-\frac{1}{\tau}\left(P_{_\textrm{++}}(E_p,s)\!-\!\overline{P}_{_\textrm{++}}(E_p,s)\right),\nonumber\\
 \!s\widetilde{P}_{\textrm{-\,-}}(E_{p},s)\!&\!+\!&\!\widetilde{\gamma}_{_-}(E_{p})\widetilde{P}_{\textrm{-\,-}}(E_{p},s)\!-\!\widetilde{\gamma}_{_-}(E_{p})P_{_\textrm{++}}(E_{p},s)\!=\nonumber\\
 \!&\!=\!&\! - \frac{1}{\tau}\left(\widetilde{P}_{\textrm{-\,-}}(E_{p},s)\!-\!\overline{\widetilde{P}}_{\textrm{-\,-}}(E_{p},s)\right).\\
\nonumber
\end{eqnarray}
Here tilde denotes the shift by $E_{_J}$ of the energy argument in
a function, {\it e.g.}
$\widetilde{P}_{\textrm{-\,-}}(E_{p},s)=P_{\textrm{-\,-}}(E_{p}\!+\!E_{_J},s)$.
The system of algebraic equations~(\ref{algeb}) can be solved for
$P_{_\textrm{++}}(E_p,s)$ and $P_{_\textrm{-\,-}}(E_p,s)$. Then,
by summing these expressions over the momenta $p$ and utilizing
Eq.~(\ref{equilP}) we obtain a closed system of equations for
qubit populations $\sigma_{_{\pm\pm}}(s)$:
\begin{eqnarray}\label{kin_nq2}
\sigma_{_\textrm{++}}(s)\!&\!=\!&\!A(s)+\frac{A(s)}{\tau}\sigma_{_\textrm{++}}(s)+\frac{B(s)}{\tau}\sigma_{\textrm{-\,-}}(s),\nonumber\\\nonumber\\
\sigma_{\textrm{-\,-}}(s)\!&\!=\!&\!C(s)+\frac{C(s)}{\tau}\sigma_{_\textrm{++}}(s)+\frac{D(s)}{\tau}\sigma_{\textrm{-\,-}}(s),
\end{eqnarray}
where the coefficients $A(s)$, $B(s)$, $C(s)$ and $D(s)$ are given
below
\begin{eqnarray}
A(s)\!&\!=\!&\!\frac{1-Z(s)}{s+1/\tau}, \mbox{ \, \ } B(s)=\frac{Z(s)e^{-E_{_J}/T}}{s+1/\tau},\nonumber\\
C(s)\!&\!=\!&\!\frac{Z(s)}{s+1/\tau}, \mbox{ \, \ }
D(s)=\frac{1-Z(s)e^{-E_{_J}/T}}{s+1/\tau}.
\end{eqnarray}
Function $Z(s)$ is
\begin{eqnarray}\label{zs0}
Z(s)=\sum_p\frac{\gamma_{_+}(E_p)\rho_{\rm{odd}}(E_p)}{s+1/\tau+\Gamma(E_p)}
\end{eqnarray}
with $\gamma_{_+}(E_p)$ and $\Gamma(E_p)$ being defined in
Eqs.~(\ref{rates3}),~(\ref{ratesnew}) and~(\ref{Gamma}),
respectively. From now on we take thermodynamic limit and replace
the sum by the integral in Eq.~(\ref{zs0}). (Thermodynamic limit
is appropriate here, since $\delta_r \ll T$.)
\begin{eqnarray}\label{zs}
Z(s)=\frac{1}{\delta_r}\int_{\Delta}^{\infty}d
E_p\nu(E_p)\frac{\gamma_{_+}(E_p)\rho_{\rm{odd}}(E_p)}{s+1/\tau+\Gamma(E_p)}.
\end{eqnarray}

The solution of Eqs.~(\ref{kin_nq2}) yields the following results
for $\sigma_{_{\pm\pm}}(s)$:
\begin{eqnarray}\label{sigma}
\sigma_{_\textrm{++}}(s)\!&\!=\!&\!\frac{1}{s}\left(1-\frac{(\tau
s+1)Z(s)}{
\tau s+Z(s)(1+e^{-E_{_J}/T})}\right), \nonumber\\\nonumber\\
\sigma_{\textrm{-\,-}}(s)\!&\!=\!&\!\frac{1}{s}\frac{(\tau
s+1)Z(s)}{\tau s+Z(s)(1+e^{-E_{_J}/T})}.
\end{eqnarray}
Equations~(\ref{sigma}) allow us to analyze the dynamics of the
qubit populations for arbitrary $\Gamma_{\textrm{in}}\tau$. Let us
point out that Eqs.~(\ref{sigma}) satisfy normalization condition
$\sigma_{_\textrm{++}}(s)+\sigma_{\textrm{-\,-}}(s)=1/s$. To find
the evolution of the populations, it is sufficient to evaluate
$\sigma_{_\textrm{++}}(t)$.

\begin{figure}
\centering
\includegraphics[height=3.0in, scale=1]{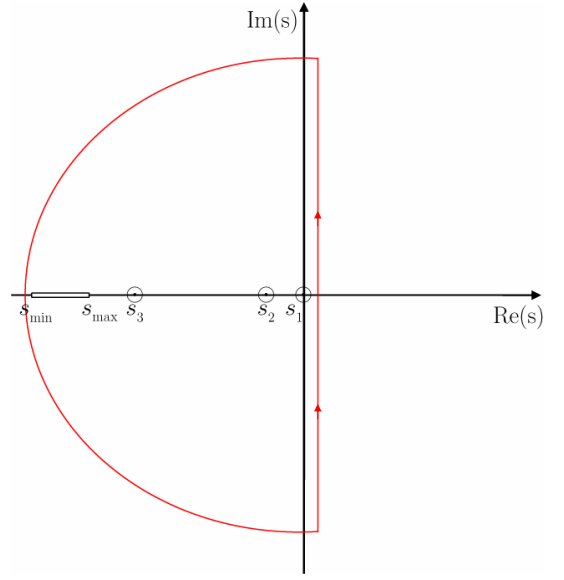}
\caption{(color online). Contour of integration (red line) chosen
to calculate inverse Laplace transform Eq.~(\ref{Brom}). Points of
nonanalytic behavior of $\sigma_{_\textrm{++}}(s)$ are shown.
(Poles at $s_1$, $s_2$, and $s_3$, and a cut $s \in
(s_{\rm{min}},s_{\rm{max}})$).} \label{fig2a}
\end{figure}

The inverse Laplace transform is given by
\begin{eqnarray}\label{Brom}
\!\sigma_{_\textrm{++}}(t)\!=\!\frac{1}{2\pi
i}\!\int_{\eta-i\infty}^{\eta+i\infty}{ds}\,\sigma_{_\textrm{++}}(s)e^{st},
\end{eqnarray}
where $\eta$ is chosen in such way that $\sigma_{_\textrm{++}}(s)$
is analytic at $\mbox{\rm{Re}} [s]>\eta$. The
integral~(\ref{Brom}) can be calculated using complex variable
calculus by closing the contour of integration as shown in
Fig.~\ref{fig2a} and analyzing the enclosed points of nonanalytic
behavior of $\sigma_{_\textrm{++}}(s)$. In general, the
singularities of $\sigma_{_\textrm{++}}(s)$ consist of three poles
and a cut. The latter is due to the singularities of the function
$Z(s)$ causing $\sigma_{_\textrm{++}}(s)$ to be nonanalytic along
the cut $s \in \left(s_{\rm min},s_{\rm max} \right)$, where
\begin{eqnarray}
s_{\rm min}\!&\!=\!&\!-\!\frac{1}{\tau}\!-\!\mbox{\rm max}
\left[\,\Gamma(E_p)\,\right]\nonumber,\\
s_{\rm max}\!&\!=\!&\!-\!\frac{1}{\tau}\!-\!\mbox{\rm min}
\left[\,\Gamma(E_p)\,\!\right]\nonumber.
\end{eqnarray}
The schematic plot of $\Gamma(E_p)$ is shown in Fig.~\ref{fig5}.

In addition to the cut, $\sigma_{_\textrm{++}}(s)$ has 3 poles.
The first one is at $s_1=0$; two more poles, $s_2$ and $s_3$, are
the solutions of the following equation in the region of
analyticity of the function $Z(s)$:
\begin{eqnarray}\label{eq_zeros}
\tau s+Z(s)\left(1+e^{-E_{_J}/T}\right)=0.
\end{eqnarray}

The preceding discussion of the analytic properties of
$\sigma_{_\textrm{++}}(s)$  is general for any ratio of the
relaxation time $\tau$ and quasiparticle escape rate
$\Gamma_{\textrm{in}}$. However, the location of the singularities
and their contribution to the integral~(\ref{Brom}) depends on
$\Gamma_{\textrm{in}}\tau$. Below we briefly present results for
two cases of interest: fast ($\Gamma_{\textrm{in}}\tau \ll 1$) and
slow ($\Gamma_{\textrm{in}}\tau \gg 1$) quasiparticle relaxation.
The detailed analysis of the singularities of
$\sigma_{_\textrm{++}}(s)$ is given in the Appendix.

In the fast relaxation regime ($ \Gamma_{\textrm{in}}\tau \ll 1$),
the contributions from the cut and  the residue at $s_3$ are small
(proportional to $\Gamma_{\textrm{in}}\tau$, see the Appendix) and
thus can be neglected. Then, relevant poles of
$\sigma_{_\textrm{++}}(s)$ in this limit are
\begin{eqnarray}\label{poles_fast}
s_1=0,\,
s_2=-\left<\!\gamma_{_+}\!\right>-\left<\!\gamma_{_-}\!\right>.
\end{eqnarray}
The integration of Eq.~(\ref{Brom}) yields, up to corrections
vanishing in the limit $\Gamma_{\textrm{in}}\tau\ll 1$,
Eqs.~(\ref{sol_sigma_tau0}) for the populations.

In the slow relaxation case, $\Gamma_{\textrm{in}}\tau \gg 1$, the
main contribution to the integral~(\ref{Brom}) comes from the cut,
and poles $s_1=0$ and $s_2$. The latter may be found by iterative
solution of Eq.~(\ref{eq_zeros}),
\[
s_2=-\frac{Z(0)}{\tau}\left(1\!+\!e^{- E_{_J}/T}\right),
\]
where function $Z(0)$ is defined in Eq.~(\ref{zs}), and in the
case of slow relaxation can be approximated as
\begin{eqnarray}\label{z_0}
Z(0)=\frac{1}{\delta_r}\int_{\Delta}^{\infty}d
E_p\nu(E_p)\frac{\gamma_{_+}(E_p)}{\Gamma(E_p)}\rho_{\rm{odd}}(E_p).
\end{eqnarray}
The residue at $s_3$ gives a smaller by factor
$1/\Gamma_{\textrm{in}}\tau\ll 1$ contribution, see the Appendix.

Taking integral in Eq.~(\ref{Brom}) along the contour enclosing
the cut shown in Fig.~\ref{fig2a}, and accounting for poles $s_1$
and $s_2$, we find
\begin{eqnarray}\label{sigma_slow}
\!\sigma_{_{++}}\!(t)\!\!&\!\!=\!\!&\!\!\frac{e^{-\!E_{_J}/T}}{1\!+\!e^{-\!
E_{_J}/T}}\!+\!\left(\!\frac{1}{1\!+\!e^{-\!
E_{_J}/T}}\!-\!Z(0)\!\right)\!\times\!\nonumber\\
\!&\!\times\!&\!\exp\!\left(\!-Z(0)\!\left(1\!+\!e^{-\!E_{_J}/T}\right)\frac{t}{\tau}\!\right)\!+\!\nonumber\\
\!&\!+\!&\!\!\frac{1}{\delta_r}\int_{\Delta}^{\infty}d
E_p\nu(E_p)\frac{\gamma_{_+}(\!E_p)\rho_{\rm{odd}}(\!E_p)}{\Gamma(E_p)}\exp\!\left(\!-\Gamma(E_p)\,t\right).\nonumber\\
\end{eqnarray}
Here we neglected the corrections to Eq.~(\ref{sigma_slow}) of the
order $1/\Gamma_{\textrm{in}}\tau$. The obtained expression for
$\!\sigma_{_{++}}\!(t)$  describes the kinetics of the qubit
populations in the slow relaxation regime. Note that the solution
Eq.~(\ref{sigma_slow})  satisfies initial conditions
$\sigma_{_{++}}(0)=1$ and is consistent with previous results.
Indeed, in the limit $\tau \! \rightarrow \! \infty$, the exponent
in the second term goes to zero and we recover Eq.~(\ref{sigma35})
(To show this one should use Eq.~(\ref{balance})).

Equation~(\ref{sigma_slow}) becomes physically transparent in the
low-temperature limit. Using the approximation~(\ref{leading}) for
$Z(0)$ and for the integral in the last term of
Eq.~(\ref{sigma_slow}) we find
\begin{eqnarray}\label{sigma_slow2}
\!\sigma_{_{++}}(t)\!&\!=\!&\!\frac{e^{- E_{_J}/T}}{1\!+\!e^{-
E_{_J}/T}}\!+\!\sqrt{\frac{T}{\pi E_{_J}}}\exp\left(\!-\frac{t}{T_1^{*}}\right)\nonumber\\
\!&\!+\!&\!\left(\frac{1}{1\!+\!e^{-
E_{_J}/T}}\!-\!\sqrt{\frac{T}{\frac{}{}\pi
E_{_J}}}\,\right)\!\exp\!\left(-\!\frac{t}{T_1}\right),
\end{eqnarray}
where relaxation times $T_1$ and $T_1^{*}$ are
\begin{eqnarray}
\frac{1}{T_1}\!=\!\frac{1}{\tau}\sqrt{\frac{T}{\pi E_{_J}}} \mbox{
and }
\frac{1}{T_1^{*}}\!=\!\frac{g_{_T}\delta_r}{4\pi}\sqrt{\frac{\omega_{_+}}{2\Delta+\omega_{_+}}}\left(1\!+\!\frac{E_{_J}}{\omega_{_+}}\right).
\end{eqnarray}

Obtained results describe the relaxation of the qubit populations
in the slow relaxation limit ($ \Gamma_{\textrm{in}}\tau \gg 1$)
discussed qualitatively in Sec.~\ref{quali}. According to
Eq.~(\ref{sigma_slow2}), in this case the process of equilibration
of qubit populations occurs in two stages. The first stage ($t\sim
T_1^{*}$) corresponds to a quasistationary state formation with
qubit populations much larger than equilibrium ones. For the
typical experimental temperatures $T\!\sim \!20$ mK, the excited
state population of the qubit is about $85 \%$. The equilibrium
populations are established in the second stage, on the time scale
of $T_1$. The relaxation time $T_1$ sets an important experimental
constraint on the frequency of repetition of qubit experiments.

We estimate now energy and phase relaxation times for the
realistic experimental parameters~\cite{Nakamura, Wallraff}:
$\Delta\!\approx\!2K$, $E_c\!\approx\!0.5K$,
$E_{_J}\!\approx\!0.3K$, $g_{_T}\!\sim\!1$, and
$T\!\approx\!20mK$. For the volume of the reservoir
$10^{-19}-10^{-17}m^3$, Eq.~(\ref{T22}) yields the phase
relaxation time $T_2\sim 10^{-5}\!-\!10^{-3}s$. Energy relaxation
depends on the relation between $\Gamma_{in}$ and $1/\tau$. Taking
$\tau\!\sim \!10^{-4}s$ and $\Gamma_{in}^{-1}\!\sim\! 10^{-5}s$,
which corresponds to the lower end of the volume range, energy
relaxation is described by Eq.~(\ref{sigma_slow2}) with
$T_1^{*}\!\sim\! 10^{-5}s$ and $T_1 \!\sim\! 10^{-3}s$.

We considered so far the effect of a single quasiparticle on the
qubit kinetics. It is possible to generalize our results onto the
case of many quasiparticles $N_{\textrm {qp}}$ residing in the
system. In this case $T_2$ becomes shorter since the quasiparticle
tunneling rate $\Gamma_{\textrm{in}}$, see Eq.~(\ref{gammain}),
should be multiplied by the number of quasiparticles in the
superconducting reservoir
\begin{equation}
\frac{1}{T_2}\simeq
\Gamma_{\textrm {in}}\cdot N_{\textrm {qp}}
\simeq\frac{g_{_\textrm{T}}n_{\textrm{qp}}}{4\pi\nu}.
\label{nqp}
\end{equation}
(Here $\nu$ is the normal density of states per unit volume and
$n_{\rm qp}$ is the density of quasiparticles in the reservoir.)
Note that the volume of the reservoir does not enter in
Eq.~(\ref{nqp}).  A finite density of quasiparticles in the
reservoir affects also the process of energy relaxation of the
qubit. The most clear example corresponds to the limit
$\Gamma_{\rm in}\tau\ll\!~\!1$ in which quasiparticle relaxation
in the reservoir occurs fast compared to the time needed for the
quasiparticle to reenter the CPB. In this limit
$T_1/T_2\sim\sqrt{E_J/T}$.

The comparison of the theoretical prediction for $T_2$ and $T_1$
with experimental data is complicated by the unknown value of
$N_{\rm qp}$ in a qubit. The quasiparticle density in a system
with a massive lead is known to deviate from the equilibrium value
in a number of experiments~\cite{Prober, Aumentado}. The estimate
of $n_{\rm qp}$ can be obtained from the kinetics of
``quasiparticle poisoning'' studied in the recent
experiments~\cite{Naaman, Schneiderman}. The observed rate of
quasiparticle entering the CPB was $10^5-10^4$ Hz. Assuming that
the bottleneck for the quasiparticles was tunneling through the
junction (rather than the diffusion in the lead), we estimate the
density of quasiparticles in the lead to be $n_{\textrm{qp}}\sim
10^{19}-10^{18}\,\textrm{m}^{-3}$. The same quasiparticle density
in a qubit with the reservoir volume $10^{-19}-10^{-17}m^3$ would
result in $N_{\textrm{qp}}\sim 1-100$.

\section{Conclusions}\label{conclusion}
We studied the kinetics of a superconducting charge qubit in the
presence of an unpaired electron. The presence of a quasiparticle
in the system leads to the decay of quantum oscillations. We
obtained master equations for the coherences and populations of
the qubit, which take into account energy exchange between the
quasiparticle and the qubit, and include the mechanisms of
quasiparticle relaxation due to electron-phonon interaction.
Finally, we found decay exponents governing the dynamics of the
qubit for different cases: fast and slow quasiparticle relaxation
in the reservoir.

We have shown that phase relaxation is determined by the
quasiparticle tunneling rate to the box $\Gamma_{\textrm{in}} \sim
g_{_T}\delta_r/4\pi$. Kinetics of the qubit populations depends on
the ratio of the quasiparticle relaxation time $\tau$ and escape
time $\Gamma_{\textrm{in}}^{-1}$. In this paper, we considered two
limits - fast ($\tau \Gamma_{\textrm{in}} \ll 1$)
 and slow ($\tau \Gamma_{\textrm{in}} \gg 1$) quasiparticle relaxation.
In the latter case, decay of qubit populations occurs in two
stages. In the first stage at $t \sim \Gamma_{\textrm{in}}^{-1}$ a
quasistationary regime is established with large nonequilibrium
excited state population. The second stage describes the
attainment of the equilibrium populations and occurs on the time
scale of $\tau\sqrt{\pi E_{_J}/T}$. In the fast relaxation case,
equilibrium qubit populations are established at $t \sim
\Gamma_{\textrm{in}}^{-1}\sqrt{\pi E_{_J}/T}$.

\begin{acknowledgments}

We thank A. Kamenev, R. Schoelkopf, P. Delsing, O.~Naaman and J.~
Aumentado  for stimulating discussions. RL would like to thank E.~
Kolomeitsev for useful comments on the manuscript. This work is
supported by NSF grants DMR 02-37296,  and DMR 04-39026.

\end{acknowledgments}

\appendix

\section{Analysis of the analytical structure of $\sigma_{_\textrm{++}}(s)$.}\label{appendix_poles}

In this appendix we study the analytic properties of
$\sigma_{_\textrm{++}}(s)$ in order to calculate the inverse
Laplace transform~(\ref{Brom}). In general, the nonanalytic
behavior of $\sigma_{_\textrm{++}}(s)$ is determined by three
poles, one of them is at $s=0$, and a cut as shown in
Fig.~\ref{fig2a}. The locations of two other poles and of the cut,
and also contributions of all the mentioned singularities in
$\sigma_{_\textrm{++}}(s)$ to the integral~(\ref{Brom}), depend on
the value of $\Gamma_{\textrm{in}}\tau$.

In the fast relaxation regime ($ \Gamma_{\textrm{in}}\tau \ll 1$),
in the vicinity of the $s=0$ pole, we find
\begin{eqnarray}\label{s1}
\sigma_{_\textrm{++}}(s)=\frac{e^{- E_{_J}/T}}{1+e^{-
E_{_J}/T}}\frac{1}{s}.
\end{eqnarray}
Two other poles $s_2$ and $s_3$ are the solutions of
Eq.~(\ref{eq_zeros}) with $s_2$ being the solution at small $s\sim
\Gamma_{\textrm{in}}$ and $s_3$ at large $s\sim 1/\tau$:
\begin{eqnarray}
s_2&=&-\left(\left<\!\gamma_{_+}\!\right>+\left<\!\gamma_{_-}\!\right>\right),\nonumber\\
s_3&=&-\frac{1}{\tau}-\mbox{\rm min}
\left[\,\Gamma(E_p)\,\right]+\frac{g_{_T}\delta_r}{4\pi}C_1.
\end{eqnarray}
Here $\Gamma(E_p)$ is defined in Eq.~(\ref{Gamma}) and $C_1$ is a
distance from the beginning of the cut in units of
$g_{_T}\delta_r/4\pi$, see Fig.~\ref{fig2a}. In the vicinity of
the second pole, $\sigma_{_\textrm{++}}(s)$ is given by
\begin{eqnarray}
\sigma_{_\textrm{++}}(s)=\left(\frac{1}{1+e^{-
E_{_J}/T}}-Z(0)\right)\frac{1}{s-s_2},
\end{eqnarray}
with $Z(0)$ defined in Eq.~(\ref{z_0}). The residue of
$\sigma_{_\textrm{++}}(s)$ at $s_3$ is proportional to
$(\tau\Gamma_{\textrm{in}})^2$. Consequently, the contribution to
the integral~(\ref{Brom}) from the pole $s_3$ is small.

In addition to the poles discussed above, nonanalyticity of
$\sigma_{_\textrm{++}}(s)$ comes from the singularities of $Z(s)$.
The function $Z(s)$ is nonanalytic along the cut $ s \in
\left(s_{\rm min},s_{\rm max} \right) $, where
\begin{eqnarray}
s_{\rm min}\!&\!=\!&\!-\frac{1}{\tau}\!-\!\mbox{\rm max}
\left[\Gamma(E_p)\right]\nonumber,\\
s_{\rm max}\!&\!=\!&\!-\!\frac{1}{\tau}\!-\!\mbox{\rm min}
\left[\Gamma(E_p)\right],\nonumber
\end{eqnarray}
with $\Gamma(E_p)$ being defined in Eq.~(\ref{Gamma}). The proper
contribution to Eq.~(\ref{Brom}) can be calculated by integrating
along the contour enclosing the cut
\begin{eqnarray}\label{cut0}
\!I_{\rm cut}\!&\!=\!&\!\frac{\!-\!1}{2\pi i}\int_{s_{\rm
min}}^{s_{\rm
max}}{ds}e^{st}\left(\sigma_{_\textrm{++}}(s\!+\!i\epsilon)\!-\!\sigma_{_\textrm{++}}(s\!-\!i\epsilon)\right).\nonumber\\
\end{eqnarray}
The discontinuity of the imaginary part of
$\sigma_{_\textrm{++}}(s)$ at the cut is
\begin{eqnarray}\label{cut1}
&\,&\sigma_{_\textrm{++}}(s\!+\!i\epsilon)\!-\!\sigma_{_\textrm{++}}(s\!-\!i\epsilon)=\nonumber\\\nonumber\\
\!&\!=\!&\!\frac{-2i\tau(\tau s\!+\!1) \mbox{\rm
Im}Z(s\!+\!i\epsilon)}{ [\tau s\!+\!\mbox{\rm
Re}Z(s)(1\!+\!e^{\!-\! E_{_J}/T})]^2\!+\![\mbox{\rm
Im}Z(\!s\!+\!i\epsilon\!)
(1\!+\!e^{\!-\! E_{_J}/T})]^2}.\nonumber\\
\end{eqnarray}
In the limit $\Gamma_{\textrm{in}}\tau \ll 1$ we find
\begin{eqnarray}\label{cut2}
\sigma_{_\textrm{++}}(s\!+\!i\epsilon)\!-\!\sigma_{_\textrm{++}}(s\!-\!i\epsilon)\!&\!\approx\!&\!\frac{-2i\tau(\tau
\Gamma_{\textrm{in}}) \mbox{\rm Im}Z(s\!+\!i\epsilon)}{
[\!-\!1\!+\!\mbox{\rm
Re}Z(s)]^2\!+\![\mbox{\rm Im}Z(\!s\!+\!i\epsilon\!)]^2},\nonumber\\\nonumber\\
\end{eqnarray}
which yields a negligible contribution to Eq.~(\ref{Brom}) from
the cut, $I_{cut} \propto \Gamma_{\textrm{in}}\tau \ll 1$.
Finally, after summing up two relevant contributions, one obtains
the result for $\sigma_{_\textrm{++}}(t)$ given in
Eq.~(\ref{sol_sigma_tau0}).

\begin{figure}
\centering
\includegraphics[width=3.1in]{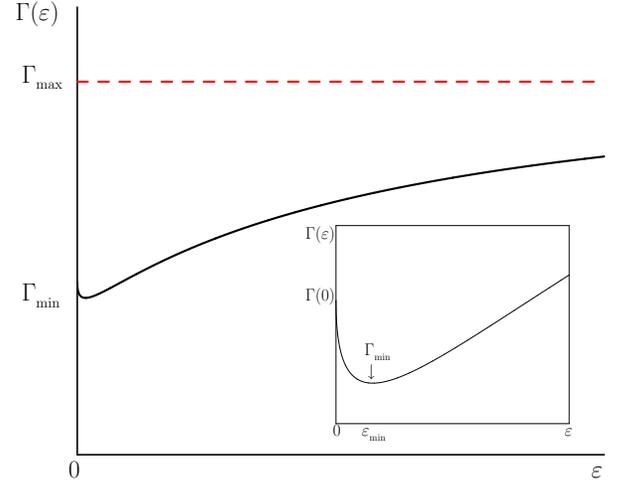}
\caption{(color online) Main panel: dependence of
$\Gamma(\varepsilon)$, defined in Eq.~(\ref{Gamma}), on
  quasiparticle energy $\varepsilon=E_p-\Delta$. Inset: dependence of
$\Gamma(\varepsilon)$ in the vicinity of $\varepsilon_{_{\rm
min}}$. The maximum value of the rate is
  $\Gamma_{\rm max}=g_{_T}\delta_r/4\pi$ (dashed line). In the limit
  $E_{_J}\ll\omega_{_+}\ll\Delta
  $, a simple analytic expression for the minimum can be found: $\varepsilon_{_{\rm
min}}=E_{_J}/25$, and
  $\frac{\Gamma_{\rm{min}}-\Gamma(0)}{\Gamma(0)}\approx-C\frac{E_{_J}}{\omega_{_+}}$, where
$\Gamma(0)=\frac{g\delta_r}{4\pi}\sqrt{\frac{\omega_{_+}}{2\Delta}}$,
and numerical constant $C \approx 0.1$.} \label{fig5}
\end{figure}

In the opposite limit of slow relaxation ($
\Gamma_{\textrm{in}}\tau \gg 1$), the first pole $s_1=0$ is the
same as in the previous case with the expression for
$\sigma_{_\textrm{++}}(s)$ given by Eq.~(\ref{s1}). The other two
poles $s_2$ and $s_3$ are found from Eq.~(\ref{eq_zeros}) assuming
$ \Gamma_{\textrm{in}}\tau \gg 1$:
\begin{eqnarray}\label{As2s3_slow}
s_2&=&-\frac{Z(0)}{\tau}\left(1\!+\!e^{- E_{_J}/T}\right),\nonumber\\
s_3&=&\!-\mbox{min}[\,\Gamma(E_p)\,]\!-\frac{1}{\tau}+\frac{C_2}{\tau^2\Gamma_{\textrm{in}}},
\end{eqnarray}
where $Z(0)$ is defined in Eq.~(\ref{z_0}), and $C_2$ is a
positive constant of the order of unity. In the vicinity of the
second pole $\sigma_{_\textrm{++}}(s)$ is given by
\begin{eqnarray}
\sigma_{_\textrm{++}}(s)=\left(\frac{1}{1+e^{-
E_{_J}/T}}-Z(0)\right)\frac{1}{s-s_2}.
\end{eqnarray}

The third pole $s_3$ lies in close proximity to the beginning of
the cut $s_{{\rm max}}$. In order to find the position of the pole
we expand the denominator of $Z(s)$ in the neighborhood of the
minimum of $\Gamma(E_p)$, see Fig.~\ref{fig5}:
\begin{eqnarray}\label{Zs3}
Z(s)\!=\!\int_{\Delta}^{\infty}{\frac{dE_p}{\delta_r}}\frac{\nu(E_p)\gamma_{_+}(E_p)\rho_{\rm{odd}}(E_p)}{s\!+\!1/\tau\!+\!\Gamma_{\rm
min}
\!+\!\Gamma''(E{_p}_{\rm min})(E_p\!-\!E{_p}_{\rm min})^2},\nonumber\\
\end{eqnarray}
and solve Eq.~(\ref{eq_zeros}) to obtain $s_3$ of
Eq.~(\ref{As2s3_slow}). Here we used the small-$y$ asymptote,
$Z(y)\propto 1/\sqrt{y}$, where $y$ is the distance from $s_{\rm
max}$ in units of $\Gamma_{\textrm{in}}$; \textit{i.e.}, we made a
substitution
$$s=-1/\tau-\Gamma_{\rm min}+\Gamma_{\textrm{in}}y$$ in Eq.~(\ref{Zs3}) and
evaluated the integral at $y \ll 1$. Then, in the neighborhood of
$s_3$ the expression for $\sigma_{_\textrm{++}}(s)$ can be written
as
\begin{eqnarray}
\sigma_{_\textrm{++}}(s)\approx\frac{\Gamma_{\textrm{in}}\tau
Z(y_0)}{\Gamma_{\textrm{in}}\tau+Z'(y_0)(1+e^{-E_{_J}/T})}\frac{1}{s-s_3},\\\nonumber
\end{eqnarray}
where $y_0$ is the corresponding solution of Eq.~(\ref{eq_zeros})
and is small: $y_0\!\sim\! (\Gamma_{\textrm{in}}\tau)^{-2} \ll 1$.
Since derivative $Z'(y_0)\propto y_0^{-3/2}$, the residue at $s_3$
is suppressed as $(\Gamma_{\textrm{in}}\tau)^{-1}$ and thus can be
neglected.

The contribution from the cut in the limit
$\Gamma_{\textrm{in}}\tau \gg 1$ can be evaluated from
Eq.~(\ref{cut1}). The discontinuity of the function
$\sigma_{_\textrm{++}}(s)$ is
\begin{eqnarray}
\sigma_{_\textrm{++}}(s\!+\!i\epsilon)\!-\!\sigma_{_\textrm{++}}(s\!-\!i\epsilon)\approx
-\frac{2i}{s}\mbox{\rm Im}Z(s\!+\!i\epsilon).
\end{eqnarray}
Hence, the contribution to Eq.~(\ref{Brom}) from the cut is
\begin{eqnarray}
\!I_{\rm cut}\!&\!=\!&\!-\!\int_{\Delta}^{\infty}\!\frac{d
E_p}{\delta_r}\nu(E_p)\!\gamma_{_+}\!(\!E_p)\!\rho_{\rm{odd}}(\!E_p)\!\int_{\!s_{\rm
min}\!}^{\!s_{\rm max}\!}{\!ds}\!\frac{e^{st}}{s}
\delta\!\left(\!s\!+\Gamma(E_p)\right)\nonumber\\
\!&\!=\!&\!\int_{\Delta}^{\infty}\frac{d
E_p}{\delta_r}\nu(E_p)\frac{\gamma_{_+}(E_p)\rho_{\rm{odd}}(E_p)}{\Gamma(E_p)}\exp\left(-\Gamma(E_p)t\right).
\end{eqnarray}
Finally, combining proper terms one finds the inverse Laplace
transform of Eq.~(\ref{sigma_slow}).


\begin{thebibliography}{99}
\bibitem{footnote} The possibility to include the Josephson tunneling
  term in $H_0$ while considering the quasiparticle tunneling
  Hamiltonian $V$ as a perturbation was demonstrated in
  Ref.~[8].
\bibitem{phonons} We do not consider here qubit relaxation originating
from the excitation of phonons by charge fluctuations across the
Josephson junction~\cite{Ioffe}. The presence of a quasiparticle
opens a relaxation channel that is faster than the mechanism
considered in Ref.~[20].
\end{thebibliography}
\end{document}